\documentclass[aps,pre,floatfix,nofootinbib
]{revtex4}
\usepackage{graphicx} \usepackage{amsmath} \usepackage{amssymb}
\usepackage[utf8]{inputenc}
\usepackage{xcolor}
\newcommand{\BEA}{\begin{eqnarray}}
\newcommand{\EEA}{\end{eqnarray}}

\newcommand{\be}{\begin{eqnarray}}
\newcommand{\ee}{\end{eqnarray}}

\newcommand{\I}{{\rm I}}
\newcommand{\II}{{\rm II}}
\renewcommand{\varrho}{R}

\newcommand{\comment}[1]{}
\newcommand{\A}{{\rm acc}}
\newcommand{\R}{{\rm rej}}
\newcommand{\F}{1}
\newcommand{\U}{0}
\renewcommand{\aa}{a_0}
\newcommand{\bb}{a_1}
\newcommand{\cc}{a_2}
\begin{document}

\title{Ultimatum game: regret or fairness? }

\author{ L.H. Aleksanyan$^{1)}$, A.E. Allahverdyan$^{1,2)}$, V.G. Bardakhchyan$^{1,2)}$}

\address{ 
$^{1)}$Alikhanian National Laboratory (Yerevan Physics Institute), Alikhanian Brothers Street 2,  Yerevan 0036, Armenia,\\
$^{2)}$Yerevan State University, 1 A. Manoogian street, Yerevan 0025, Armenia
}

\begin{abstract}
In the ultimatum game, the challenge is to explain why responders reject non-zero offers thereby defying classical rationality. Fairness and related notions have been the main explanations so far. We explain this rejection behavior via the following principle: if the responder regrets less about losing the offer than the proposer regrets not offering the best option, the offer is rejected. This principle qualifies as a rational punishing behavior and it replaces the experimentally falsified classical rationality (the subgame perfect Nash equilibrium) that leads to accepting any non-zero offer. The principle is implemented via the transitive regret theory for probabilistic lotteries. The expected utility implementation is a limiting case of this.
We show that several experimental results normally prescribed to fairness and intent-recognition can be given an alternative explanation via rational punishment; e.g. the comparison between ``fair'' and ``superfair'', the behavior under raising the stakes {\it etc}. Hence we also propose experiments that can distinguish these two scenarios (fairness versus regret-based punishment). They assume different utilities for the proposer and responder. We focus on the mini-ultimatum version of the game and also show how it can emerge from a more general setup of the game.



\end{abstract}

\maketitle

\section{Introduction}
\label{intro}

The ultimatum game is an important experiment in behavioral economics and game theory that has been used to study human decision-making and social behavior \cite{guth}. In this game, two players are given a certain amount of money $A$, and one player, the proposer $\I$, is asked to divide the money between $\I$ and the other player $\II$ (the responder). $\II$ decides whether to accept or reject the offer. If $\II$ accepts the offer, both players keep their respective shares of the money. However, if $\II$ rejects the offer, both players receive nothing. The ultimatum game has been used in a variety of research contexts, including economics, psychology, neuroscience, and sociology, to investigate questions such as how people make decisions, and how gender, individual differences, and social norms affect these decisions \cite{solnick, solnick2,hessel, chuah, jones}.

The traditional assumption in the ultimatum game is that people are rational and self-interested in the sense of sub-game perfect Nash equilibria. Hence $\II$ will accept whatever is given. Knowing about that $\I$ should offer the smallest possible amount of money. This is the only (sub-game perfect, i.e. stable) Nash equilibrium of the game. However, in practice, proposers often offer more than the minimum, and responders often reject offers that are too small; e.g. proposers offered around 30-40\% of $A$, and responders tended to reject offers that were below 20-25\% of $A$ \cite{guth}. 

Any non-zero offer is better than nothing. Why do responders reject them? 

One of the main explanations for the rejection behavior is that people care about fairness whenever this serves their own interests \cite{rand,ruffle,straub,fehr}. This explanation assumes that when responders and proposers interchange their roles, former responders need not be willing to offer a fair amount \cite{fehr}. One problem with {\it a priori} assumptions of fairness is that its general quantification is unclear. Indeed, there are experimental results showing that the simplest definition of fairness (i.e. equal split) is unstable with respect to small perturbations \cite{guthfairisunfair}. This may mean that the concept of fairness is not employed by people in a sensible way.  
On the other hand, there are experimental results showing that an individual demonstrating rejection behavior in the ultimatum game 
need not behave prosocially in other games, e.g. need not cooperate in the prisoner's dilemma game \cite{japan_rejection}. 
Alternative existing theories for explaining the rejection behavior in the ultimatum game are reviewed in Appendix \ref{review}.

Our study posits that rejection can be a rational punishment in the long run if $\II$ is trying to force $\I$ to provide better options.
Once rejecting means that $\II$ sacrifices short-term rationality, there is a requirement for long-term rationality: the rejection is meaningful only when $\II$ regrets about losing the offered (smallest) amount less than $\I$ regrets about not offering a larger amount. Here, regret is understood as referring to the generalized expected utility theory for comparing probabilistic lotteries developed in \cite{savage1951,bell,loomes_sugden}, and recently advanced in \cite{we_regret}.  
Ref.~\cite{we_regret} demonstrated that regret theory can be reconciled with basic principles of rationality, such as stochastic dominance and transitivity. Ref.~\cite{we_regret} also shows that the regret theory can solve the classical difficulties of the expected utility theory. In particular, it can solve Allais' paradox, because the regret theory generally does not hold the sure thing principle \cite{we_regret}, and it can solve Savage's omelet paradox since the regret theory is a counter-factual mean of comparing two (or more) probabilistic lotteries \cite{we_regret}. Once regret is a rational mean of comparing agent's different actions with each other, it differs conceptually from envy, which is concerned with things beyond one's control. 


We show that several experimental results in the ultimatum game that are usually attributed to fairness can be given an alternative interpretation via regret-based rational punishment. This includes the following effects. {\it (i)} The very definition of fair division, which for equal utilities of $\I$ and $\II$ amounts to dividing the initial sum $A$ over two halves. {\it (ii)} Comparison between unfair/fair and unfair/superfair offers. {\it (iii)} Behavior changes (or lack thereof) under increasing stakes, i.e. proportionally increasing both the overall divided sum $A$ and the magnitudes of offers. In all these cases, regret-based punishment and fairness predict similar results only for identical utilities of $\I$ and $\II$. Their predictions are markedly different for different utilities. The mechanism behind this difference is that the richer agent can endure more losses and hence demand more, whereas according to the fairness theory, the richer agent will likely continue promoting fairness. At the very least, this is the case when the fairness is supplemented with intention recognition, as was suggested for explaining {\it (ii)} via fairness \cite{falk}. 


The rest of this paper is organized as follows. The next section reminds the ultimatum game and its Nash equilibrium. Section \ref{geno} reformulates the ultimatum game in terms of probabilistic lotteries and shows how to apply the regret theory. The ultimatum game with two outcomes (mini-ultimatum game) is studied in section \ref{two}. Here we study in detail the similarities between assuming fairness and rational punishment. Section \ref{three} studies the ultimatum game for three or more offers and discusses in which sense the optimal three-offer game is reduced to the optimal two-offer game. We summarize our results in the last section.

\section{The ultimatum game and its Nash equilibrium}
\label{trado}

We specify the rules of the ultimatum game for two offers.
This is sometimes called the mini-ultimatum game \cite{bolton_zwick,population,gale,falk}. This assumption is sufficient for the traditional solution. In our regret-based solution, the number of offers can be arbitrary, but the two outcomes are the minimal situation to start from. There are two players $\I$ and $\II$ and a certain amount of money $A$. Now $\I$ offers to $\II$ either $A-\aa$ (action $\U$) or $A-\bb$ (action $\F$):
\BEA
\label{gomesh}
A>\aa>\bb.
\EEA
$\II$ can accept any of these offers or reject them. Money is lost in the latter case: neither $\I$ nor $\II$ receives anything. In more detail, we can design this as a sequential game \cite{myerson}, where 
$\II$ has three options: $(\A,\A)$ (accept if $\F$ or $\U$), $(\A,\R)$ (accept if $\F$, reject if $\U$), and $(\R,\R)$ (reject if $\F$ or $\U$). The outcomes can be written as follows:
\BEA
\label{bad}
\begin{tabular}{||c|c|c|c||}
  \hline
~$ {\I}/{\II}$~  & $(\A,\A)$ & $(\A,\R)$ & $(\R,\R)$  \\
  \hline
  ~$\F$~ & ~$\bb, \, A-\bb$~ & ~${\bb, \, A-\bb}$~ & ~$0, \, 0$~ \\
 \hline\hline
  ~$\U$~ & ~${\aa, \, A-\aa}$~ & ~$0, \, 0$~ & ~$0, \, 0$~ \\
  \hline
\end{tabular}
\label{ta}
\EEA
This table makes clear that there are two Nash equilibrium points here: $[\U, (\A,\A)]$ and $[\F, (\A,\R)]$. However, $[\F, (\A,\R)]$ is not subgame perfect \cite{myerson}, i.e. acting $\R$ in response to $\U$ means that $\II$ looses money. Put differently, by declaring $(\A,\R)$, $\II$ makes a non-credible threat. Thus, people who act $(\A,\R)$ (as it frequently happens in experiments \cite{guth}) do not follow the 
subgame perfect Nash equilibrium \cite{myerson}. 

\section{The ultimatum game reformulated via regret}
\label{geno}

\subsection{Lotteries and utilities}

We start with a reformulation of the ultimatum game, where the number of divisions (offers) by $\I$ is finite and equals $n+1$: within each offer $\I$ keeps $a_k$ \$ and gives $A-a_k$ \$ to $\II$:
\BEA
A>a_0> ... > a_n>0.
\label{lopes}
\EEA
Each $a_k$ in (\ref{lopes}) is offered by $\I$ with prior probability $\pi_k\geq 0$: $\sum_{i=0}^n \pi_i=1$. 

$\II$ will have two options: either accept whatever is given or accept the offer $A-a_k$ with (conditional) probability $p_k$:
\BEA
p_0 \leq p_1 \leq ... \leq p_n = 1 .
\label{vega}
\EEA
The ordering in (\ref{vega}) is a consequence of (\ref{lopes}); the last condition $p_n=1$ means that the best possible option for $\II$ is never rejected. We emphasize that this assumption of always accepting $a_n$ is non-trivial.
We shall denote:
\BEA
\hat{a}\equiv\{a_i\}_{i=0}^n, \quad \hat{p}\equiv\{p_i\}_{i=0}^n,\quad \hat{\pi}\equiv\{\pi_i\}_{i=0}^n.
\EEA
Hence $\II$ chooses between two lotteries: 
\BEA
\label{olo1}
\II_{\rm acc}= \begin{pmatrix}
A-a_0 & A-a_1 & ... & A-a_n \\
\pi_0 & \pi_1 & ... & \pi_n
\end{pmatrix}, \qquad
\II_{\rm rej}= \begin{pmatrix}
A-a_0 & A-a_1 & ... & A-a_n & 0 \\
p_0\pi_0 & p_1\pi_1 & ... & p_n\pi_n & 1-\sum_{i=0}^{n} p_i \pi_i
\end{pmatrix}.
\EEA
Note that the outcomes in $\II_{\rm acc}$ and $\II_{\rm rej}$ are independent of each other. 

$\I$ faces a choice between the following lotteries:
\BEA
\I_{n}= \begin{pmatrix}
a_n \\
1
\end{pmatrix} \qquad
\I_{0}= \begin{pmatrix}
a_0 & a_1 & ... & a_n & 0 \\
p_0\pi_0 & p_1\pi_1 & ... & p_n\pi_n & 1-\sum_{i=0}^{n} p_i \pi_i
\end{pmatrix},
\label{olo2}
\EEA
where $\I_n$ means offering $\II$ the best possible (for $\II$) outcome, while $\I_0$ relates to $\II_{\rm rej}$ in (\ref{olo1}).

Each monetary outcome in (\ref{olo1}, \ref{olo2}) has its utility $u(x)$ which is  a non-decreasing function of $x$. We assume $u(0)=0$ without loss of generality. We assume that both $\I$ and $\II$ have the same utility function; this assumption will be revised later. 

\subsection{Regret and its calculation}

We start with $\rho_\II (\II_{\rm acc};\II_{\rm rej}\to A-a_k)$ which is the regret experienced by $\II$ about not choosing $\II_{\rm acc}$, provided that $\II_{\rm rej}$ was chosen and the outcome $A-a_k$ has been realized. This quantity is defined as follows \cite{we_regret}: 
\BEA
\rho_\II(\II_{\rm acc};\II_{\rm rej}\to A-a_k)=\sum_{l=0}^n \pi_l f[u(A-a_l)-u(A-a_k)],\\
\rho_\II(\II_{\rm acc};\II_{\rm rej}\to 0)=\sum_{l=0}^n \pi_l f[u(A-a_l)].
\label{koni}
\EEA
Indeed, since it is not known which outcome would have been realized within $\II_{\rm acc}$ all its outcomes enter into 
(\ref{koni}) with their probabilities. The function $f[x]$ in (\ref{koni}) compares different outcomes; hence it naturally holds \cite{we_regret}: 
\BEA
\label{sign}
f(x\geq 0)\geq 0, \qquad f(x\leq 0)\leq 0, \qquad f(0)=0.
\EEA
Generally, $f[x]$ accounts for both regret and appreciation. We get a pure regret (appreciation) if $f[x\leq 0]=0$ ($f[x\geq 0]=0$). 

Mean regret of not choosing $\II_{\rm acc}$ thus reads
\BEA
\rho_\II(\II_{\rm acc};\II_{\rm rej})=\sum_{k=0}^n p_k\pi_k
\rho_\II(\II_{\rm acc};\II_{\rm rej}\to A-a_k) +\Big(1-\sum_{l=0}^n p_l\pi_l \Big) \rho_\II(\II_{\rm acc};\II_{\rm rej}\to 0).
\label{koni2}
\EEA
Now $\rho_\II(\II_{\rm rej};\II_{\rm acc})$ is calculated in the same way assuming that $\II_{\rm acc}$ was taken {\it etc}. 
To decide which one, $\II_{\rm rej}$ or $\II_{\rm acc}$ was eventually worth taking, we look at the difference \cite{we_regret}: 
\BEA
\varrho_\II&=&\rho_\II(\II_{\rm acc};\II_{\rm rej})-\rho_\II(\II_{\rm rej};\II_{\rm acc})\\
&=& \Big(1-\sum_{i=0}^{n}p_i \pi_i\Big) \sum_{k=0}^{n}
\pi_k g[u(A-a_k)] + \sum_{0\leq i<j\leq n} (p_j-p_i)\pi_i \pi_j g[u(A-a_i)-u(A-a_j)],
\label{koku} 
\EEA
where we denoted [cf.~(\ref{sign})]:
\BEA
\label{r8}
&&g[x]\equiv f[x]-f[-x],\\
\label{u2}
&& g[x] = - g[-x],\\
&& g[x]\geq g[y]\quad {\rm for}\quad x\geq y.
\label{uu2}
\EEA
$g[x]$ is the main regret function we shall work with. 

Likewise, we calculate from (\ref{olo2}) for $\I$:
\BEA
\varrho_{\I}= \rho_{\I}(\I_n; \I_{0})-\rho_{\I}(\I_{0}; \I_{n})=
\sum_{i=0}^{n} p_i \pi_i g[u(a_n)-u(a_i)]+\Big(1-\sum_{i=0}^{n}p_i \pi_i\Big)g[u(a_n)],
\label{hop}
\EEA
where we assumed that $\I$ and $\II$ have the same regret function $g[x]$ in (\ref{sh}). 

\subsection{The solution}

Note from (\ref{olo1}) that every positive outcome in $\II_{\rm acc}$ have a larger probability than the same outcome in $\II_{\rm rej}$, i.e. . 
the probabilities in $\II_{\rm acc}$ stochastically dominate those in $\II_{\rm rej}$. This leads from (\ref{koku}--\ref{uu2}) in \cite{we_regret} to: 
\BEA
\varrho_\II>0,
\label{ceka}
\EEA
i.e. when taking $\II_{\rm acc}$ one rejects less than taking $\II_{\rm rej}$: the short-term rationality dictates choosing $\II_{\rm acc}$.  
$\II$ decides to sacrifice this by acting $\II_{\rm rej}$. However, long-term rationality still demands that $\I$ regrets more for not offering $a_n$, than $\II$ regrets for rejecting. Only under this condition, $\II$ can hope that the rejection will force $\I$ to change the offer because only under this condition the rejection by $\II$ will have a chance to be a sustainable action. This situation can be described as 
follows:
\BEA
\label{mrr}
\II {\rm ~wins~over~} \I ~{\rm if}~0<\varrho_{\II}<\varrho_{\I},\\
\I {\rm ~wins~over~} \II ~{\rm if}~\varrho_{\II}>\varrho_{\I}.
\label{grr}
\EEA

\subsection{Regret function}

The simplest choice of the regret function is 
\BEA
\label{bj}
g[x]\propto x, 
\EEA
and (\ref{koku}--\ref{hop}) reduce to difference between expected utilities ${\rm EU}$ of separate lotteries, e.g.
\BEA
\varrho_{\II}={\rm EU}(\II_{\rm acc})-{\rm EU}(\II_{\rm rej})=
\sum_{l=0}^n\pi_lu(A-a_l)-\sum_{l=0}^n\pi_lp_lu(A-a_l).
\EEA
We shall see below the expected utility version of the theory is simple and can produce useful results, though it is not always sufficiently non-trivial; cf.~Appendix \ref{expected}.

Modulo an affine transformation, $f[x]\to af[x]+b$ ($a>0$) of $f[x]$, there is only one regret function $g(x)$ which ensures that pairwise comparison between lotteries holds transitivity \cite{we_regret}:
\BEA
g[x]=\sinh \left[{x}/{\beta}\right],\quad \beta>0,
\label{sh}
\EEA
where $\beta>0$ is a parameter. We revert to (\ref{bj}) for $\beta\to\infty$. The
intuitive meaning of (\ref{sh}) is explained in Appendix \ref{sidor}.

\section{Mini-ultimatum game: two offers}
\label{two}

We focus on the mini-ultimatum case $n=1$, where there are two offers only \cite{bolton_zwick,population,gale}. 
Now $p_0<1$, $p_1=1$, $\pi_0+\pi_1=1$, while in the last sum of (\ref{koku}) only the term with $i=0$ and $j=1$ survives. Condition (\ref{mrr}) (i.e. $\II$ wins) amounts from (\ref{koku}, \ref{hop}) to 
\BEA
&&\pi_0\left(g[u(A-\aa)]-g[u(A-\bb)]+g[u(A-\bb)-u(A-\aa)]    \right)\nonumber\\
<
&&g[u(\bb)]+\frac{p_0}{1-p_0}g[u(\bb)-u(\aa)]-g[u(A-\bb)]
+g[u(A-\bb)-u(A-\aa)].
\label{gra}
\EEA
For the expected utility situation (\ref{bj}) we get from (\ref{gra}) 
\BEA
\label{ono}
u(a_1)-u(A-a_0)>\frac{p_0}{1-p_0}(u(a_0)-u(a_1)).
\EEA
It is seen that (\ref{ono}) holds $u(a_1)>u(A-a_0)$, i.e. $a_1+a_0>A$ taking into account that $u(x)$ is monotonic function. If this condition holds, $p_0$ can be taken sufficiently small so that (\ref{ono}) holds. Thus $\II$ wins for $a_1+a_0>A$, while $\I$ wins for $a_1+a_0<A$. Now $\pi_0$ equals either $0$ or $1$. 

Let us turn to the more general case (\ref{sh}). 
Employing (\ref{sh}) and standard relations
\BEA
\label{gloss}
&\sinh x-\sinh y=2\sinh\frac{x-y}{2}\cosh\frac{x+y}{2},\quad 
\sinh x=2\sinh\frac{x}{2}\cosh\frac{x}{2},\\
&\cosh x-\cosh y=2\sinh\frac{x-y}{2}\sinh\frac{x+y}{2},
\EEA
we find for (\ref{gra}):
\BEA
&\kappa\equiv -g[u(A-\aa)]+g[u(A-\bb)]-g[u(A-\bb)-u(A-\aa)] \nonumber\\
&=4\sinh\frac{u(A-\bb)-u(A-\aa)}{2\beta}\sinh\frac{u(A-\aa)}{2\beta}
\sinh\frac{u(A-\bb)}{2\beta}
\geq 0.
\label{gl}
\EEA
Recalling that $\aa>\bb$, while $u(x)$ and $g(x)$ are monotonously increasing, we see that Eqs.~(\ref{gra}--\ref{gl}) lead to two possibilities analyzed below. 

\subsection{$\II$ wins over $\I$}

If in (\ref{gra}) we have
\BEA
0< g[u(\bb)]-g[u(A-\bb)]+g[u(A-\bb)-u(A-\aa)],
\label{gras2}
\EEA
then
$\II$ can choose the probability $p=p_0$ from 
\BEA
p_0< \frac{g[u(\bb)]-g[u(A-\bb)]+g[u(A-\bb)-u(A-\aa)]}{g[u(\bb)]-g[u(A-\bb)]+g[u(A-\bb)-u(A-\aa)]+g[u(\aa)-u(\bb)]}<1,
\label{kuk}
\EEA
so that $\I$ regrets more for any choice of $1\geq \pi_0\geq 0$, i.e. (\ref{gra}) holds. Thus, under condition (\ref{gras2}), $\II$ can force $\I$ into proposing $\bb$. Note that the actual value of $p$ is not important provided that it holds (\ref{kuk}) because $\I$ is now expected to offer $\bb$ with probability $1$, i.e. $\II$ has nothing to reject. As follows from (\ref{lopes}) and the monotonicity of $g[u(x)]$ as a function of $x$, condition (\ref{gras2}) does hold for 
\BEA
\bb>A/2.
\label{pitt}
\EEA
The meaning of (\ref{pitt}) is that even in the worst case, $\I$ does not offer less than the half of the money. And $\I$ is punished for this, because now $\II$ can ensure that for any choice of $\pi_0$, the regret of $\I$ is larger provided that (\ref{kuk}) is satisfied. 

\subsection{$\I$ wins over $\II$}

For $\bb<A/2$ it is possible that (\ref{gras2}) is inverted:
\BEA
0> g[u(\bb)]-g[u(A-\bb)]+g[u(A-\bb)-u(A-\aa)].
\label{gras4}
\EEA
Now $\I$ can make $\pi_0$ sufficiently small and invert (\ref{gra}) for any $p_0$. To this end, it is needed that [cf.~(\ref{gl})]
\be
\pi_0<\pi_c=1+\frac{g[u(A-a_0)]-g[u(\bb)]}{\kappa}. 
\label{gras5}
\ee
Thus, if (\ref{gras5}) holds, then $\II$ is not able to force $\I$ to regretting the action of not giving the best option $\bb$ to $\II$. Then $\II$ will just accept whatever is given. The situation is now in equilibrium since $\I$ is well-motivated to act with probability $\pi_0$ close but smaller than $\pi_c$. Using $\pi_0>\pi_c$ means that $\II$ will force $\I$ into regret, while $\pi_0$ substantially lower than $\pi_c$ means that $\I$ looses utility.

Note from (\ref{gras5}) that $\pi_c\geq 1$ for
\BEA
g[u(\bb)]\leq g[u(A-\aa)], ~~{\rm i.e.}~~ \bb+\aa\leq A.
\label{gora}
\EEA
Under the last condition in (\ref{gora}), $\II$ cannot force $\I$ into regret and $\I$ always employs the best option $a$ (i.e. $\pi=1<\pi_c$), where the outcome for $\I$ is the largest one. Taken together with $\bb<A/2$, condition $\bb+\aa>A$ has a transparent meaning: If the amount to which $\II$ agrees with probability one is too large (e.g. $\bb\to 0$), $\I$ can punish $\II$ by getting almost the whole money, i.e. $\aa\to A$ under (\ref{gora}). 

\comment{The average utilities found by $\I$ and $\II$ will read, respectively:
\be
\label{gras6}
&& \bar{u}_{\I}=\pi u(\aa)+(1-\pi)u(\bb),\\
&& \bar{u}_{\II}=\pi u(A-\aa)+(1-\pi)u(A-\bb).
\label{gras7}
\ee }

\begin{figure}[!ht]
\centering
    \includegraphics[width=0.7\columnwidth]{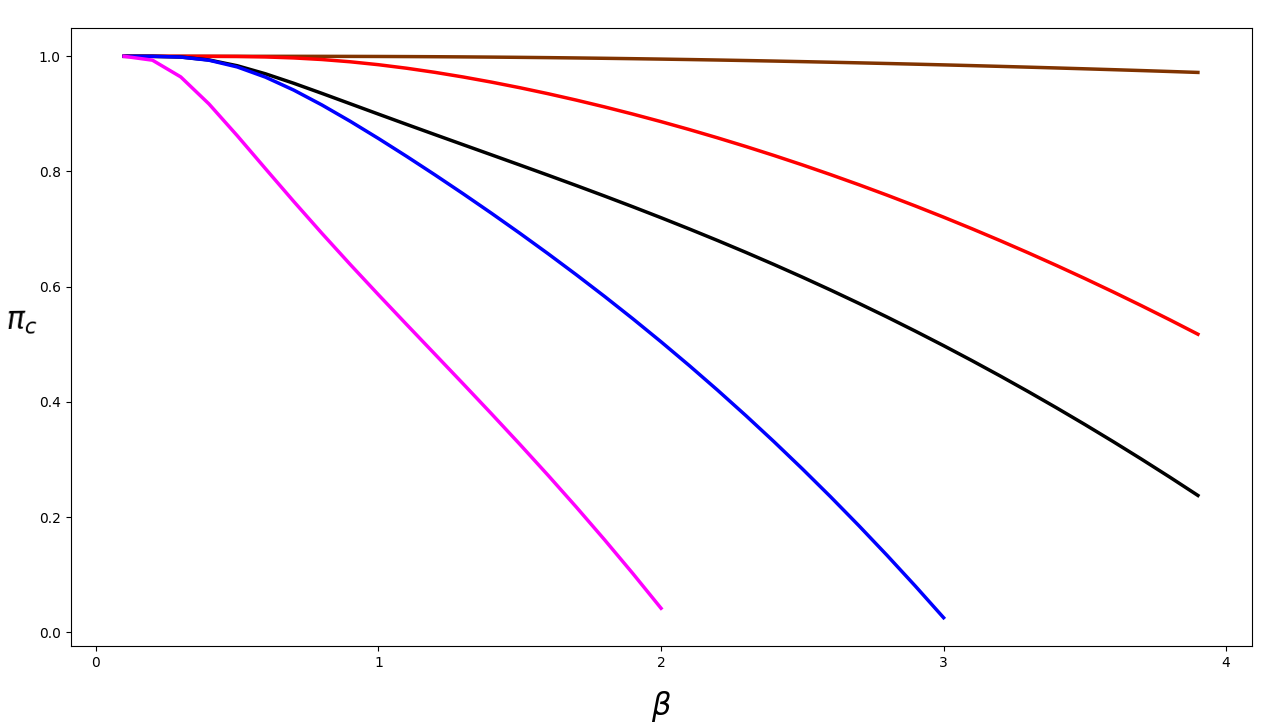}
    \caption{ The critical probability $\pi_c$ (given by (\ref{gras5})) versus $\beta$ for $A=10$ and $u(x)=x$ (linear utility). Parameters of the curves are as follows (from top to bottom):
    brown $a_0=8.1$, $a_1=2$; red $a_0=8.1$, $a_1=3$; black $a_0=7.1$, $a_1=4$; blue $a_0=8.1$, $a_1=4$; magenta $a_0=8.1$, $a_1=4.5$.
    $\pi_c$ is larger (stays closer to $1$) when $a_1$ decreases for a fixed $a_0$, or $a_0$ decreases for a fixed $a_1$. Whenever $a_1$ approaches $0.5$, $\pi_c$ quickly tends to $0$ as a function of $\beta$. 
    }
\label{figpi}
\end{figure}

\subsection{Fairness {\it versus} regret for different utilities}

We did not impose any fairness assumptions into the above solution of the ultimatum game, because we focus on the rational punishment behavior. Nevertheless, some of our results are similar to those obtained by assuming fairness; see also the next subsection.
Indeed, we saw that for $A>a_0>a_1>A/2$, $\II$ wins over $\I$, while for $a_1<A/2$ the winning strategies by $\I$ do exist. Now $A/2$ plays a special role here because it is assumed that utilities are equal for $\I$ and $\II$. Let us take them differently, and assume the well-known logarithmic utilities for $\I$ and $\II$: 
\be
\label{ex}
u_k(x)=\ln\left(\frac{x}{\gamma_k}+1\right),\qquad k=\I,~\II,
\ee
where $\gamma_k>0$ are positive parameters that characterize the agent; i.e. $\gamma_k>0$ defines the threshold of the concave (risk-averse) utility $u_k(x)$ ($u(0)=0$), because only for $\frac{x}{\gamma_k}\ll 1$ we have $u_k(x)\simeq 0$. Hence, $\gamma_k$ defines the initial wealth, since only for $\frac{x}{\gamma_k}\gtrsim 1$ the decision maker will care about money.
From the viewpoint of (\ref{ex}), only the ratio $x/\gamma_k$ matters, i.e. it does not matter whether one decreases stakes $x$, or increases the initial wealth $\gamma_k$. 

Now above formulas trivially modify for different utilities, e.g.  $g[u(\aa)]\to g[u_\I(\aa)]$ and $g[u(A-\aa)]\to g[u_\II(A-\aa)]$. Instead of (\ref{pitt}) and (\ref{gora}) we have, respectively,
\BEA
\bb>\frac{A}{1+(\gamma_\II/\gamma_\I)}, \qquad \bb>\frac{\gamma_\I}{\gamma_\II}(A-\aa).
\label{bala}
\EEA
Eqs.~(\ref{bala}) show that the richer agent endures more losses and hence is able to win over the poorer agent under a wider range of conditions; e.g. two time richer $\II$ ($\frac{\gamma_\II}{\gamma_\I}=2$) wins under $\bb>A/3$ according to the first equation in (\ref{bala}). We believe that the honest application of the fairness assumption will result in the opposite result, e.g. richer $\II$ will request less not more. 

\subsection{Unfair/fair {\it versus} unfair/superfair}

Ref.~\cite{falk} reports on an interesting experimental effect [see also \cite{guth}], which the authors of Ref.~\cite{falk} interpret in terms of intent-regarding behavior of $\II$. In our notations the effect is described as follows: responders $\II$ tend to reject more in the situation $a_0>\frac{A}{2}=a_1$, then for $a_0>\frac{A}{2}$ and $a_1=A-a_0<\frac{A}{2}$. Put differently, $\II$'s reject the unfair offer $a_0$ more when it comes with the fair one $a_1=A/2$, than when it comes with the superfair one $a_1=A-a_0$. 

This result is compatible with the above theory. Indeed, (\ref{gras5}) predicts that for the unfair/fair situation $a_0>\frac{A}{2}=a_1$ we get $\pi_c<0$, which means that $\II$ wins, i.e. $\II$ will reject hoping to force $\I$ towards the fair outcome $a_1=A/2$. Likewise, for the unfair/superfair situation $A=a_0+a_1$ we get $\pi_c=1$ [see (\ref{gras5})], i.e. $\I$ wins and the rejection is meaningless. 

The authors of Ref.~\cite{falk} explain the effect as follows: since $\II$ cares also about the intentions of $\I$, $\II$ will recognize that for the unfair/superfair situation $\I$ is not really unfair, since offering the superfair option implies too much generosity, and hence $\II$ will reject less. Our explanation seems to us more practical at least from the normative viewpoint: $\II$ does not reject in the unfair/superfair situation simply because there is no hope to force $\I$ towards the superfair option. 

It is easy to envisage a situation, where the explanations based on (resp.) regret and intent-regarding will lead to different outcomes. Consider the case of different utilities and deliberately assume in (\ref{ex}) that $\II$ has more initial wealth than $\I$: $\gamma_\II>\gamma_\I$. Now we again implement the above comparison between unfair/ fair ($a_0>\frac{A}{2}=a_1$) and unfair/superfair ($a_1=A-a_0<A/2$). If the intent-regarding explanation is correct, $\II$ will tend to reject the unfair/superfair offer even less (or at least the same) than for $\gamma_\II=\gamma_\I$, because now $\I$ is poorer and should be less motivated to make the wasteful superfair offer for $\II$. In contrast, the explanation based on regret tells that now $\I$ will get more rejections from $\II$, simply because the latter is less susceptible to losing money. Indeed, using (\ref{gras5}, \ref{ex}) with $\gamma_\II>\gamma_\I$ (i.e. $u_\I(x)>u_\II(x)$) and $a_1=A-a_0<A/2$ we get [cf.~(\ref{gl})]: 
\BEA
\pi_c=1-\frac{g[u_\I(a_1)]- g[u_\II(A-a_0)]}{{-g[u_\II(A-\aa)]+g[u_\II(A-\bb)]-g[u_\II(A-\bb)-u_\II(A-\aa)]}}<1.
\EEA
Note that the unfair/ fair ($a_0>\frac{A}{2}=a_1$) offer is naturally always rejected for $\gamma_\II>\gamma_\I$.

\subsection{Changing of the stakes and the influence of $\beta$}
\label{prado}

The domain of probabilities, where $\I$ wins over $\II$ is given by (\ref{gras5}). How does the domain change when all stakes are increased proportionally, i.e. $A$, $a_0$, and $a_1$ are multiplied by a constant factor (say $10$)? In posing this question we for clarity assume the linear utility $u(x)=x$. 

Recall that $\beta$ controls deviations from the effective expected utility theory, larger $\beta$ meaning smaller deviations. Increasing the stakes for a fixed $\beta$, is equivalent to lowering $\beta$; cf.~(\ref{sh}). Hence, the same question can be asked differently: how this domain is influenced by the parameter $\beta$ in the (\ref{sh})? 

\begin{figure}[!ht]
\centering    \includegraphics[width=0.8\columnwidth]{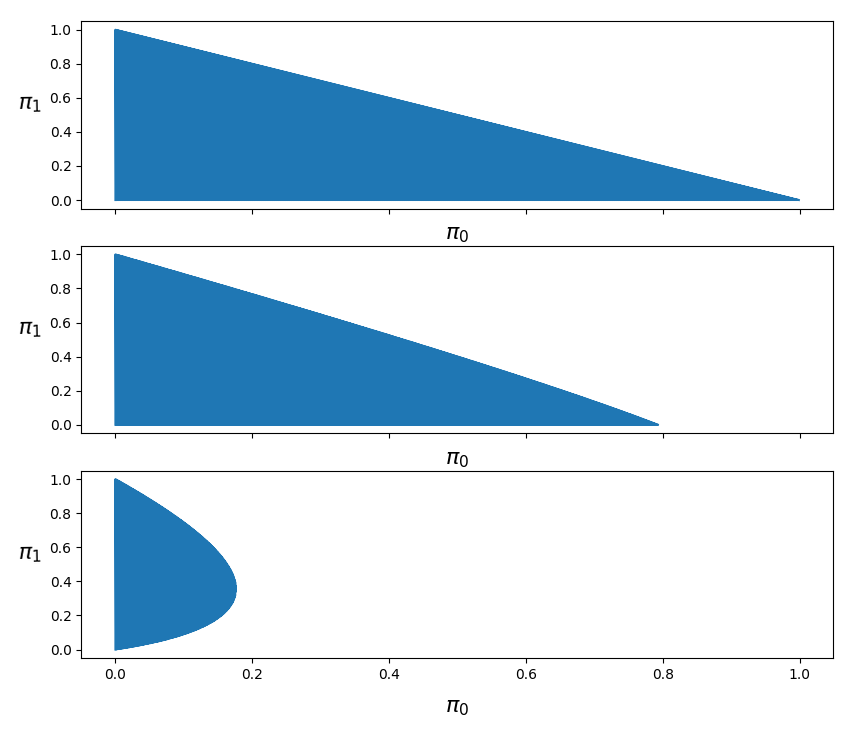}
\caption{For three offers $\aa = 90, \bb = 60, \cc = 40$ [cf.~(\ref{lopes})], this figure shows the values of prior probabilities $\pi_0$ and $\pi_1$ (depicted by dark domains), where the proposer $\I$ wins getting smaller regret than the responder $\II$. From top to bottom: $\beta=1$, $\beta=10$ and $\beta=30$.}
\label{bettas}
\end{figure}

Fig.~\ref{figpi} shows that $\pi_c$ is a monotonously decaying function of $\beta$ for $A/2>a_1$. This means that raising the stakes (i.e. lowering $\beta$) improves the situation for $\I$. The same conclusion is reached for the three-offer ultimatum game; see Fig.~\ref{bettas}.  
Note that the question of raising stakes was studied experimentally in \cite{stakes}, where it was concluded that the experimental data on the offer by $\I$ did not show a significant deviation from the proportional increase upon raising $A$ in the full ultimatum game. Within our approach, this can correspond to the expected utility limit $\beta\to\infty$, where the winning domain for $\I$ indeed does not depend on $\beta$. Indeed, (\ref{ono}) shows that for $u(x)=x$ (and $\beta\to\infty$) the winning conditions do not change upon multiplying $A$, $a_0$ and $a_1$ over a positive constant factor. 

To our knowledge, raising the stakes was not addressed for the mini-ultimatum game. However, the existing theories on the cost of fairness indicate that the position of $\I$ is going to improve upon raising the stake \cite{stakes,telser}. According to the cost of fairness explanation, for larger stakes people do not want fairness, they want money, i.e. tend to accept relatively small, but now absolutely large offers. We see again that our results reproduce certain effects on fairness without employing this notion. 

\subsection{The optimal mean utility of proposer $\I$}

We saw in (\ref{gras4}, \ref{gras5}) that for $\bb<A/2$, $\I$ should employ $\pi_0<{\rm min}[\pi_c,1]$ in order to win. 
We now assume that $a_1<A/2$ is fixed (this is the offer to which $\II$ agrees certainly), while $a_0$ is chosen by $\I$. One possibility to choose $a_0$ optimally is to look at the maximum mean utility:
\BEA
\label{ola}
{\cal U}_\I={\rm max}_{a_0}
\Big[\pi_0a_0+(1-\pi_0)a_1 \Big| \pi_0<{\rm min}[\pi_c,1]    \Big],
\EEA
which is maximized over $a_0$ under condition $\pi_0<{\rm min}[\pi_c,1]$. 
Note from that (\ref{gras5}, \ref{gl}) that $\pi_c>1$ for $a_0\leq A-a_1$. Hence the maximization in 
(\ref{ola}) can be written as
\BEA
\label{ola2}
{\cal U}_\I={\rm max}_{a_0\geq A-a_1}\Big[\pi_c (a_0-a_1)\Big]+a_1,
\EEA
where we recall that the maximization is carried out for a fixed $a_1<A/2$. 
Numerical study of (\ref{ola2}) shows that for a sufficiently large $\beta$, where we are qualitatively close to the expected utility regime, the maximum in (\ref{ola2}) is reached for $a_0= A-a_1$, which leads to ${\cal U}_\II=A-a_1$. In contrast, for smaller $\beta$ the maximization in (\ref{ola2}) is reached for $a_0> A-a_1$, i.e. ${\cal U}_\II>A-a_1$. Here are some numbers confirming this. For $A=\beta=1$ we found  $a_0= A-a_1$ and ${\cal U}_\II=A-a_1$ for all values of $a_1<A/2$. For $A=1$ and $\beta=0.1$ we got: ${\cal U}_\II=0.523509$ and $a_0= 0.701289$ for $a_1=0.49$; ${\cal U}_\II=0.797194$ and $a_0=0.907669$ for $a_1=0.4$; 
${\cal U}_\II=0.877306$ and $a_0=0.942645$ for $a_1=0.35$. 

\section{Three (and more) offers}
\label{three}

Here we work out (\ref{koku}, \ref{hop}) and (\ref{ceka}--\ref{grr}) for $n\geq 2$, i.e. three or more offers. Define 
the set $\Omega_n$ of winning actions for $\I$, such that for each $\hat\pi\in\Omega_n$
\BEA
\label{n7}
\Delta R[\hat{a},\hat{p},\hat{\pi}]\equiv R_{\I}[\hat{a},\hat{p},\hat{\pi}]-R_{\II}[\hat{a},\hat{p},\hat{\pi}]<0~~{\rm for~any}~\hat p,
\EEA
where $\hat p$ and $\hat a$ are defined in (\ref{vega}, \ref{lopes}), respectively. Equivalently, we can define $\Omega_n$ via
\BEA
\label{konrad}
{\rm max}_{\hat p}\Big(  \Delta R[\hat{a},\hat{p},\hat{\pi}]\Big)<0,
\EEA
where the maximization in (\ref{konrad}) goes over all $\hat p$ in (\ref{vega}) for a given $\hat a$ in (\ref{lopes}). Since $\Delta R[\hat{a},\hat{p},\hat{\pi}]$ in (\ref{konrad}) is linear over $\hat p$, and the allowed domain (\ref{vega}) of $\hat p$ is convex, the maximum in (\ref{konrad}) is reached at the end-point of the allowed domain of $\hat p$, i.e. we need to check (\ref{konrad}) only for non-trivial end-points
\BEA
\label{tur}
\hat p^{[1]}=(0,1,1,...,1),~~\hat p^{[2]}= (0,0,1,...,1), ..., ~~\hat p^{[n]}=(0,0,0,...,1).
\EEA
Hence $\Omega_n$ is determined from (\ref{konrad}, \ref{tur}) via simultaneous validity of $n$ inequalities:
\BEA
\label{konrad2}
\Delta R[\hat{a},\hat{p}^{[k]},\hat{\pi}]<0,\quad k=1,...,n. 
\EEA
Fig.~\ref{bettas} reports numerical results for the wining domain (\ref{konrad2}) of $\I$ in the three-outcome situation $n=2$.  
Note that the winning domains of $\I$ shrink upon increasing $\beta$. The same effect was discussed in section \ref{prado} for $n=1$ (two offers). 

The numerical comparison between scenarios involving two and three offers reveals the following effect of intermediate offers.  
To illustrate this effect, consider the two-offer situation with $A=100$, $a_0=96$, $a_1=0$ and $a_2=45$ ($\beta=10$ and $u(x)=x$ for concreteness). Here the domain $\Omega_1$ of winning actions for $\I$ is empty, i.e. $\II$ wins over $\I$. However, when adding an intermediate offer, i.e. considering $a_0=96$, $a_1=60$ and $a_2=45$ (with the same $\beta=10$ and $A=100$), there emerges a non-empty set $\Omega_2$. Likewise, an intermediate offer can improve the maximal mean outcome for $\I$. For example, 
\BEA
\label{but1}
{\rm max}_{(\pi_0,\pi_1,\pi_2)\in\Omega_2}\Big[ \pi_0\times 96+\pi_1\times 60+\pi_2\times 43  \Big] =61.01,\\
{\rm max}_{(\pi_0,\pi_2)\in\Omega_1}\Big[ \pi_0\times 96+\pi_1\times 0+\pi_2\times 43  \Big]=56.25.
\label{but2}
\EEA
Now ${\rm max}_{(\pi_0,\pi_1,\pi_2)\in\Omega_2}\Big[ \pi_0 a_0+\pi_1 a_1+\pi_2 a_2  \Big]$ cannot decrease when going from the two-offer situation to the three-offer situation, but there are cases when this quantity does not change upon adding e.g. an intermediate offer. For instance,
\BEA
\label{but3}
&&{\rm max}_{(\pi_0,\pi_1,\pi_2)\in\Omega_2}\Big[ \pi_0\times 90+\pi_1\times 60+\pi_2\times 40  \Big] \\
&&={{\rm max}}_{ {(\pi_0,\pi_2)\in\Omega_1}} \Big[ \pi_0\times 90+\pi_1\times 0+\pi_2\times 40  \Big]=79.5.
\label{but4}
\EEA
Both examples (\ref{but1}, \ref{but2}) and (\ref{but3}, \ref{but4}) are possible for the expected utility regret $g[x]\propto x$, as explicitly shown in Appendix \ref{expected}.

There is a sense in which the three-offer ultimatum game approximately reduces to the two-offer situation. We checked numerically that the following relation holds
\BEA
\label{vt1}
&& {\cal U}_\I^{[1]}\equiv {\rm max}_{(\pi_1,\pi_2)\in\Omega_1,\, a_1}\Big[ u(a_1)\pi_1+u(a_2)\pi_2  \Big]\\
\leq && {\cal U}_\I^{[2]}\equiv {\rm max}_{(\pi_0,\pi_1,\pi_2)\in\Omega_2,\, a_0>a_1}\Big[ u(a_0)\pi_0+u(a_1)\pi_1+u(a_2)\pi_2  \Big].
\label{vt2}
\EEA
Eq.~(\ref{vt2}) refers to maximization over the winning strategies of $\I$ in the three offer situation, and simultaneously over the two offers $a_0$ and $a_1$ [holding $a_0>a_1$ from (\ref{lopes})]. These are the offers to which $\II$ need not agree with probability $1$; cf.~(\ref{vega}). The rationale of (\ref{vt2}) is that $a_2$ is fixed, since it is determined by $\II$ as the offer $\II$ will accept certainly, while the remaining two offers $a_0$ and $a_1$ are determined by $\I$ from maximizing the mean utility. Eq.~(\ref{vt1}) is the same as (\ref{vt2}), but restricted to the two-offer situation. Variables $a_0$ and $a_1$ in (\ref{vt2}) respect $a_0>a_1>a_2$ [cf.~(\ref{lopes})], where $a_2$ is a fixed parameter. Likewise, variable $a_1$ in (\ref{vt1}) respect $a_1>a_2$, where $a_2$ is the same parameter as in (\ref{vt2}). 

The equality in (\ref{vt2}) is realized for the expected utility regret $g[x]=x$; see Appendix \ref{expected}. The inequality in (\ref{vt2}) means that the three-offer situation is better for $\I$ than the two-offer situation, i.e. the mini-ultimatum game. However, for realistic choices of parameters, the difference ${\cal U}_\I^{[2]}-{\cal U}_\I^{[1]}$ is sufficiently small; see Table~\ref{baba}. 

\comment{
The results of this analysis are as follows: in some cases with two offers (\(\aa, \cc\)), 
it was observed that the proposer's  advantage in terms of minimizing regret over the responder was absent. However, with the introduction of an intermediary offer denoted as \(\bb\), a range of values (\(\pi_0, \pi_1\)) 
emerged where a transition occurred, leading to the proposer gaining an advantageous position over the responder. An illustrative instance (Fig. 1) involves (\(\aa = 90\)), \((\bb = 60\)), and \((\cc = 40\)), with varying \(\beta\) values. Remarkably, when \((\aa = 90\)) and \((\cc = 40\)), no feasible solutions were ascertainable that would result in a victory for the responder.

It is interesting to note that the expression \(max(\aa * \pi_0 + \bb * \pi_1 + \cc * \pi_2)\) = 79.5 in found (\(\pi_0, \pi_1\)) space (\(for  \beta = 10, \aa = 90, \bb = 60, \cc = 40\)), which means that responder didn't mind to accept even very small offers.
}

\begin{table}
\label{baba}

\begin{tabular}{||c||c|c|c|c|c||}
  \hline
~$ {(\xi, [\pi])}/{(\aa, \bb, \cc, \beta)}$~ & $(59, 51, 47, 17)$ & $(70, 54, 46, 16)$ & $(69, 55, 47, 15)$ &$(77, 61, 41, 18)$&$(72, 56, 43, 18)$ \\
  \hline\hline
  ~${\cal U}_\I^{[1]}$~ & $53.24$ & $55.36$ & $54.04$ & $62.6$ & $59.24$ \\
 \hline
  ~$(\pi_0,\pi_2)$~ & $(0.52, 0.48)$ & $(0.39, 0.61)$ & $(0.32, 0.68)$ & $(0.6, 0.4)$ & $(0.56, 0.34)$ \\
  \hline\hline
  ~$\widetilde{{\cal U}}_\I^{[2]}$~ & $53.44$ & $55.6$ & $54.3$ & $62.8$ & $59.46$ \\
  \hline
  ~$(\pi_0,\pi_1,\pi_2)$~ & $(0.39, 0.44, 0.17)$ & $(0.27, 0.39, 0.34)$ & $(0.23, 0.28, 0.49)$ &$(0.55, 0.1, 0.35)$ & $(0.46, 0.24, 0.3)$ \\
  \hline
\end{tabular}
\label{ta}
\caption{Numerical values for quantities defined around (\ref{vt1}, \ref{vt2}) for $A=100$. Instead of the maximization given in (\ref{vt2}), we defined $\widetilde{{\cal U}}_\I^{[2]}\equiv {\rm max}_{(\pi_0,\pi_1,\pi_2)\in\Omega_2,\,a_1}\Big[ u(a_0)\pi_0+u(a_1)\pi_1+u(a_2)\pi_2  \Big]$ as follows: first we calculate the optimal $a_2$ from (\ref{vt1}), and then employ it in $\widetilde{{\cal U}}_\I^{[2]}$, which is optimized over $a_1$. For the presented parameter values we get $\widetilde{{\cal U}}_\I^{[2]}< {{\cal U}}_\I^{[2]}$, but still $\widetilde{{\cal U}}_\I^{[2]}\approx {{\cal U}}_\I^{[2]}$. The third and fifth rows denote the optimal values for (resp.) $(\pi_0,\pi_1)$ and $(\pi_0,\pi_1,\pi_2)$ obtained from maximizing that leads to (resp.) ${{\cal U}}_\I^{[1]}$ and $\widetilde{{\cal U}}_\I^{[2]}$.
It is seen that $\widetilde{{\cal U}}_\I^{[2]}> {{\cal U}}_\I^{[1]}$ and $\widetilde{{\cal U}}_\I^{[2]}\approx {{\cal U}}_\I^{[1]}$. 
}

\end{table}

\section{Summary}

The ultimatum game demonstrates a discrepancy between traditional rationality, defined by the subgame perfect Nash equilibrium, and real-world experimental outcomes. Contrary to the classical rational expectations, responders frequently reject sufficiently small offers (compared to the total sum $A$ to be divided), while proposers (possibly anticipating this rejection) do not offer small amounts.

To explain this rejection behavior, we propose a principle: ultimatum offers are more likely to be declined when the responder experiences less regret about rejecting the offer than the proposer would feel for not presenting a more favorable offer. In other words, when the responder is more at ease with the idea of losing the deal, while the proposer would regret deeper for not proposing a better offer, rejections become more likely. Though regret can be applied to quantifying that feeling as well, it is not simply an emotional feeling. It is an objective means of comparing two probabilistic lotteries that holds several crucial principles of rationality and thereby generalizes the expected utility theory that can be also regarded as a limiting form of regret \cite{we_regret}. 

We demonstrate that various observed phenomena in the ultimatum game, typically attributed to fairness, can alternatively be understood through the lens of rational punishment quantified via regret. These include:

1. Fair Division: In the standard concept of fair division, where participants $\I$ and $\II$ have equal utility, the initial sum $A$ is divided into two equal halves. Some of our findings align with those derived from fairness-based assumptions, e.g. the responder $\II$ wins if all offers by the proposer $\I$ are larger than $A/2$, while viable winning strategies for the proposer $\I$ exists when at least one offer is smaller than $A/2$.

2. Responses to unfair/fair {\it versus} unfair/superfair offers. Our results suggest that $\II$ rejects the unfair offer more in the unfair/fair scenario than in the unfair/superfair situation. There is no hope of steering $\I$ toward the superfair option, which explains this effect.

3. Behavior changes with increasing the stakes: we investigated how behavior changes (or remains constant) as both the total sum $A$ and the offer magnitudes increase proportionally. Our explanations are contingent upon whether the utilities for $\I$ and $\II$ are identical, revealing distinct predictions when these functions differ.
This distinction arises because wealthier participants can tolerate more significant losses and, thus, demand a larger share, contrary to what the fairness theory suggests. This demonstrates a case where considerations of fairness are supplemented by the need to recognize participants' intentions, particularly in scenarios with varying utility values.

Lastly, we show that when $\I$ optimizes both offers and strategies to maximize mean utility, the presence of three offers gives no substantial advantages over a simpler mini-ultimatum game with only two offers. Unfortunately, we lack a theoretical argument to explain why this equivalence holds, but extensive numerical testing across diverse parameter values confirms its validity.

To summarize, the regret theory provides a psychologically realistic framework for understanding human decision-making. It considers not only material gains but also the emotional consequences of choices, aligning well with observed human behavior, particularly in situations involving fairness. However, quantifying regret is subjective and varies among individuals, posing a challenge for generalization to e.g.
the multiple-respondent ultimatum game \cite{hegemon}.

\bibliographystyle{IEEEtran} 
\bibliography{ultimatum}

\appendix

\section{Alternative theories explaining rejection behavior in the ultimatum game}
\label{review}

In addition to the fairness theory reviewed in section \ref{intro} several other explanations were proposed for the rejection behavior in the ultimatum game. According to the anonymity hypothesis, the rejection behavior is caused by players seeking to please (or establish a long-term relationship) with the experimenter \cite{bolton_zwick}. Proposers wish to show that they are not greedy, hence they do not propose a minimal amount, while responders would like to seem not in need of money, i.e. they reject small offers. The hypothesis posed serious methodological questions that have been met in practice \cite{bolton_zwick}, i.e. the hypothesis -- which assumes that in reality, people would behave according to the classical scenario -- has been falsified \cite{bolton_zwick}. On the other hand, the anonymity hypothesis explains non-trivial outcomes of the dictator game, where the decision of the responder does not have any influence on the outcomes \cite{bolton_zwick}.
The anonymity hypothesis relates to the desire to be perceived as fair, which was also studied in the context of the 
dictator game \cite{fairfair}. 

Another hypothesis states that the rejection behavior is evolutionary stable given that the ultimatum game is played in populations of agents and holds certain assumptions \cite{population}. The main of them is that people interchange their roles as proposers and responders, while their behavior as responders is a slow variable \cite{population}. However, when applied to the basic situation of two players, the scenario of Ref.~\cite{population} leads to an untenable conclusion that the rejection behavior emerges independently of the parameters of the offers.

Yet another explanation of the rejection behavior focused on the role of emotions experienced by responders \cite{pillutla}. In particular, responders can be envious, which may explain higher rates of rejection of small offers in the ultimatum game \cite{georg, loyola, kato}. Envy occurs when a player cares about not only their own payoff but also about the relative payoff between them and the other player; e.g. the envy model described in \cite{georg} amounts to choosing a specific utility function that depends on the monetary outcomes of both players. It is important to note, however, that envy is an irrational feeling that cannot be advised from a normative perspective. 

\section{The expected utility choice for the regret function}
\label{expected}

For the expected utility choice (\ref{bj}) of the regret function we have
\BEA
&& \varrho_{\II}={\rm EU}(\II_{\rm acc})-{\rm EU}(\II_{\rm rej})=
\sum_{l=0}^n\pi_lu(A-a_l)-\sum_{l=0}^n\pi_lp_lu(A-a_l),\\
&&\varrho_{\II}=\rho_{\II}(\II_{\rm acc}; \II_{\rm rej})-\rho_{\II}(\II_{\rm rej}; \II_{\rm acc})
={\rm EU}(\II_{\rm acc})-{\rm EU}(\II_{\rm rej}),\\
&&{\rm EU}(\II_{\rm acc})=\sum_{l=0}^n\pi_lu(A-a_l),\qquad {\rm EU}(\II_{\rm rej})=\sum_{l=0}^n\pi_lp_lu(A-a_l),\\
&&\varrho_{\I}=\rho_{\I}(\I_{n}; \I_{0})-\rho_{\I}(\I_{0}; \I_{n})
={\rm EU}(\I_{n})-{\rm EU}(\I_{0}),\qquad
{\rm EU}(\I_{n})=u(a_n),\qquad {\rm EU}(\I_{0})=\sum_{l=0}^n\pi_lp_lu(a_l),\\
&&\varrho_{\II}-\varrho_{\I}=\sum_{l=0}^{n-1}\pi_l\Big[p_lu(a_l)+(1-p_l)u(A-a_l)-u(a_n)\Big].
\label{eu}
\EEA  

Now $\II$ want to make $\varrho_{\II}-\varrho_{\I}$ in (\ref{eu}) possibly small. To this end, $\II$ chooses:
\BEA
\label{gur}
p_l=0~~{\rm for}~~ u(a_l)>u(A-a_l)~~{\rm or}~~a_l>A/2,\\
p_l=1~~{\rm for}~~ u(a_l)<u(A-a_l)~~{\rm or}~~a_l<A/2.
\label{man}
\EEA
Thus the winning domain $\{\pi_k\}_{k=0}^n$ of $\I$ for given $(a_0>...>a_n)$ is determined by putting (\ref{gur}, \ref{man}) into (\ref{eu})
\BEA
\varrho_{\II}-\varrho_{\I}=\sum_{a_l>A/2}\pi_l\Big[u(A-a_l)-u(a_n)\Big] +\sum_{a_l<A/2,\, a_l\not=a_n}\pi_l\Big[u(a_l)-u(a_n)\Big]>0.
\label{eu2}
\EEA

Let us work out (\ref{eu2}) for $n=2$ (three offers). Now winning domains for $\I$ are found from
\BEA
\label{eu3}
&&\varrho_{\II}-\varrho_{\I}=\pi_0[u(A-a_0)-u(a_2)]+\pi_1[u(A-a_1)-u(a_2)]>0 ~~{\rm for}~~a_0>a_1>\frac{A}{2}>a_2, \\
&&\varrho_{\II}-\varrho_{\I}=\pi_0[u(A-a_0)-u(a_2)]+\pi_1[u(a_1)-u(a_2)]>0 ~~{\rm for}~~a_0>\frac{A}{2}>a_1>a_2.
\label{eu4}
\EEA
In each regime (\ref{eu3}) and (\ref{eu4}) we set to maximize the mean utility for $\I$:
\BEA
\label{kamp}
{\cal U}_\I={\rm max}_{\hat \pi}\Big[\sum_{k=0}^2\pi_ku(a_k)\Big]={\rm max}_{\hat \pi}\Big[\pi_0[u(a_0)-u(a_2)]+\pi_1[u(a_1)-u(a_2)]\Big]+u(a_2),
\EEA
where the maximization is carried out over $(\pi_0, \pi_1)$ that hold (\ref{eu3}) or (\ref{eu4}).

Within the regime (\ref{eu3}) we are back to the two-offer situation, i.e. for $u(A-a_0)>u(a_2)$ (or $a_0+a_2<A$) the maximization in (\ref{kamp}) produces $\pi_0=1$, while for $u(A-a_0)<u(a_2)g$, (\ref{eu3}) is invalid, i.e. $\II$ wins. Now for (\ref{eu4}) the maximization in (\ref{kamp}) is reached at the border, i.e. for $\varrho_{\II}-\varrho_{\I}\to 0+$ and leads to
\BEA
{\cal U}_\I= \pi_c[u(a_0)-u(a_2)]+(1-\pi_c) [u(a_1)-u(a_2)]+u(a_2),\qquad 
\pi_c=1-{\rm Max}\Big[ 0, \frac{u(a_2)-u(A-a_0)}{u(a_1)-u(A-a_0)} \Big].
\label{kats}
\EEA
The non-trivial fact about (\ref{kats}) is that for $u(A-a_0)\geq u(a_2)$ the action of $\I$ is probabilistic and amounts to offering $A-a_0$ and $A-a_1$ with probabilities $\pi_c$ and $1-\pi_c$, respectively. The meaning of this probabilistic action was already discussed in the main text: from the two-offer viewpoint the regime $u(A-a_0)<u(a_2)$ makes $\II$ to win over $\I$, and then employing the intermediate offer $a_1$ (for the present case it should hold $a_1<A/2$, as (\ref{eu4}) shows) leads to $\I$ winning over $\II$. 

We checked that if (\ref{kats}) is maximized (in its validity regime, i.e. for $a_1<A/2$ and $a_0+a_2>A$) over $a_0$, then the maximum of ${\cal U}_\I$ is reached for $a_0+a_2\to A$, i.e. for $\pi_c\to 1$. In that sense, we are effectively back to the two-offer situation. 

\section{Transitive regret}
\label{sidor}

The intuitive meaning of (\ref{sh}) is explained as follows \cite{we_regret}: the exponential
form $f[x]\propto e^{\beta x}$ is the only one that holds the multiplicativity feature $f[x+y]=f[x]f[y]$ that produces 
(\ref{sh}) via (\ref{r8}). For certain applications, the transitivity can be lifted and more general forms of $f[x]$ can be employed \cite{we_regret}, but we do not focus on such forms here. Eq.~(\ref{sh}) shows that we are back from the regret theory to the expected utility theory for $\beta\to\infty$; cf.~(\ref{bj}).

We note that $g[x]$ in (\ref{sh}) holds the following two features that make a general sense in the intuitive context of regret:
\BEA
\label{ne}
g[x+y]> g[x]+g[y],\qquad x>0,~~y>0,\\
g[x-y]\leq g[x]-g[y],\qquad x>y>0,
\label{okhi}
\EEA
where (\ref{okhi}) is deduced in (\ref{gl}), and (\ref{ne}) is derived along the same lines. Eqs.~(\ref{ne}, \ref{okhi}) make intuitive sense.
According to (\ref{ne}), regret tends to be larger when the regret-inducing stimulus is presented in total $x+y$ rather than in separate $x$ and $y$ pieces. Likewise, (\ref{okhi}) means that the regret is larger when two stimuli of different signs are arranged in parallel, as compared to the total, smaller, and positive stimulus $x-y$.

\end{document}